\documentclass[aps,twocolumn,preprintnumbers,showpacs,showkeys,nofootinbib%
]{revtex4}
\usepackage{epsfig}
\usepackage{amssymb,amsmath,amsfonts,amsthm,graphicx,psfrag}
\graphicspath{{./Figures/}}                 
\usepackage{array}
\usepackage{picture}
\usepackage{subfig}
\usepackage{graphicx}
\usepackage{textpos}

\newcommand{\noi}{\noindent}
\newcommand{\beq}{\begin{equation}}
\newcommand{\eeq}{\end{equation}}
\newcommand{\bea}{\begin{eqnarray}}
\newcommand{\eea}{\end{eqnarray}}

\newcommand{\Tab}[1]{Table~\ref{#1}}

\newcommand{\tr}{\operatorname{Tr}}

\newcommand{\bc}{{\it bc~}}
\newcommand{\fc}{{\it fc~}}

\newcommand{\ds}{\displaystyle}

\newcommand{\aasymt}{{\cal A}}

\newcommand*\xbar[1]{%
  \hbox{%
    \vbox{%
      \hrule height 0.5pt 
      \kern0.5ex
      \hbox{%
        \kern-0.1em
        \ensuremath{#1}%
        \kern-0.1em
      }%
    }%
  }%
}

\newcommand{\aasym}{\langle \Delta_{A^2} \rangle}

\begin{document}

\title{The $\big\langle A^2 \big\rangle$ asymmetry and propagators
in lattice $SU(2)$ gluodynamics at $T>T_c$.}

\author{V.~G.~Bornyakov}
\affiliation{Institute for High Energy Physics NRC "Kurchatov Institute", 142281 Protvino, Russia \\
School of Biomedicine, Far East Federal University, 690950 Vladivostok, Russia \\
Institute of Theoretical and Experimental Physics NRC "Kurchatov Institute", 117259 Moscow, Russia}

\author{V.~K.~Mitrjushkin}
\affiliation{Joint Institute for Nuclear Research, 141980 Dubna, Russia \\
and Institute of Theoretical and Experimental Physics NRC "Kurchatov Institute", 117259 Moscow, Russia}

\author{R.~N.~Rogalyov}
\affiliation{Institute for High Energy Physics NRC "Kurchatov Institute", 142281 Protvino, Russia \\
}


\begin{abstract}
We study numerically the chromoelectric-chromomagnetic
asymmetry of the dimension two $A^2$ gluon condensate
as well as the {\it transverse and longitudinal}
gluon propagators at $T>T_c$ in the Landau-gauge $SU(2)$ lattice
gauge theory with a particular emphasis on finite-volume effects.
We show that previously found so called symmetric point at which asymmetry changes sign
is an artifact of the finite volume effects. We find that with increasing temperature
the asymmetry decreases approaching zero value from above in agreement with perturbative result.
Instead of the asymmetry we suggest the ratio of the
transverse to longitudinal propagator taken at zero momentum as an indicator of  the boundary of 
the postconfinement domain and find it at $T \simeq 1.7 T_c$.

\end{abstract}



\maketitle

\section{Introduction}
\label{sec:introduction}

Studies of the dimension-two gauge-boson condensate
\beq
\big\langle A^2 \big\rangle = g^2 \big\langle  A^a_\mu (x)  A^a_\mu (x) \big\rangle \;
\eeq
in the last 15 years were initiated
by the Ref.~\cite{Gubarev:2000eu},
where it was shown that the nonperturbative part of $\big\langle A^2 \big\rangle$
is completely determined by contribution of the topological defects (monopoles)
responsible for confinement in the compact electrodynamics. Since monopole condensation is one of
the most popular scenarios of confinement also in nonabelian gauge theories, this observation
suggested that the gluon dimension-two condensate plays
an important role in the studies of infrared properties
of Yang--Mills theories as well.

In spite of some earlier considerations of the composite operator
$A^2(x) = A_\mu^a(x) A_\mu^a(x) $ (\cite{Lavelle:1988eg} {\it etc}),
for long time it was disregarded in the OPE approach
because of its gauge dependence.

In the Landau gauge, the operator $A^2(x)$
is BRST invariant (on mass shell) and multiplicatively
renormalizable, as was shown in \cite{Verschelde:2001ia,Dudal:2003np} in the
$\overline {MS}$ scheme.
Later it was argued \cite{Slavnov:2004rz} that the matrix element
$\big\langle A^2 \big\rangle$ is gauge-invariant in spite of
gauge dependence of the respective operator, still
it does not appear in the expansions of products of
gauge-invariant composite operators \cite{Bykov:2005tx}.

The effective potential for $\big\langle A^2 \big\rangle$ was obtained
in \cite{Verschelde:2001ia,Browne:2003uv} indicating
nonvanishing value of $\big\langle A^2 \big\rangle$
and thus dynamical gluon mass generation.

In the OPE approach, $\big\langle A^2 \big\rangle$ was used
for the parametrization of soft nonperturbative contributions
to the Green functions, for a review see \cite{Boucaud:2011ug}.
Thus it was extracted from their high momentum
behavior \cite{Blossier:2013oma, Boucaud:2013jwa}.

It was shown that, over the momentum range $2.5\div 7$~GeV,
ghost and gluon propagators evaluated on a lattice agree with
the respective perturbative estimates only when corrections due
to $\big\langle A^2 \big\rangle$ condensate are taken into
account \cite{Boucaud:2000nd, Boucaud:2008gn}.

In a series of papers (see e.g. \cite{Boucaud:2011ug,Pene:2011kg}
and references therein) $\big\langle A^2 \big\rangle$ was computed numerically
from fits to lattice data for the gluon and ghost propagators as well as 3-gluon
and ghost-gluon vertices.
For example, in a specific MOM-type renormalization scheme defined by a zero incoming ghost
momentum\footnote{Also referred to as the Taylor scheme}
($\mu=10$~GeV), the following values 
for the $N_f=2+1+1$ QCD were found \cite{Blossier:2013ioa}:\\[1mm]  
\hspace*{4mm}$\big\langle A^2 \big\rangle = 2.8(8)\mbox{~GeV}^2 $  \qquad
(OPE up to $\ds {1\over p^4}$)  \\[1mm]
\hspace*{4mm} $\big\langle A^2 \big\rangle = 3.8(6)\mbox{~GeV}^2 $ \qquad
(OPE up to $\ds {1\over p^6}$)\;  \\[1mm]
in order to obtain the QCD coupling constant
$\alpha_{\overline{MS}}(M_Z) = 0.1198(4)(8)(6)\; .$

The $\big\langle A^2 \big\rangle$ condensate was also intensively studied
in the refined Gribov-Zwanziger (RGZ) approach
\cite{Dudal:2008sp,Dudal:2010tf,Cucchieri:2011ig}.
Other studies of this condensate include
\cite{Kondo:2001nq,Dudal:2002pq,RuizArriola:2004en}.

In Ref.~\cite{Gubarev:2000eu}
the $\big\langle A^2 \big\rangle$ was related to
the confinement-deconfinement transition in $4D$ compact $U(1)$ gauge theory.
In this theory the confinement and deconfinement phases are separated by the
phase transition at zero temperature. It was found that the nonperturbative
part of the condensate drops at critical coupling. This observation
raised hopes that the $\big\langle A^2 \big\rangle$ condensate might be also
of relevance for the finite temperature transition in the $4D$ non-Abelian 
theories.

There are two $A^2$ condensates at nonzero temperature, electric $\big\langle A_E^2 \big\rangle$ and
magnetic $\big\langle A_M^2 \big\rangle$:
\bea
\big\langle A_E^2 \big\rangle &=& g^2 \big\langle  A^a_4 (x)  A^a_4 (x) \big\rangle, \\ \nonumber
\big\langle A_M^2 \big\rangle &=& g^2 \big\langle  A^a_i (x)  A^a_i (x) \big\rangle. \nonumber
\eea

The quantity of particular interest is the (color) electric-magnetic 
asymmetry introduced
in \cite{Chernodub:2008kf}:
\beq
 \aasym \equiv \big\langle A_E^2 \big\rangle - \frac{1}{3} \big\langle A_M^2 \big\rangle \; .
\eeq
Later we will also use the dimensionless quantity
\beq
\aasymt = {\big\langle \Delta_{A^2}(x) \big\rangle \over T^2} \; .
\eeq

Within the OPE approach and in the $p_4=0$ approximation,
it was shown \cite{Chakraborty:2012kx}
that the asymmetry contributes to the quark propagator at nonzero temperatures.

The main interest in the asymmetry stems from
its possible relation to both the confinement-deconfinement
transition and dynamics in the deconfinement phase.

There is some range of temperature values just above $T_c$, where the quark-gluon plasma (QGP)
demonstrates special properties and cannot be treated perturbatively
see e.g. \cite{Karsch:2001cy}. In particular
the QGP pressure differs from the ideal-gas value.
In the $SU(3)$ theory, an indication of this range is also provided by the behavior
of the renormalized Polyakov loop, which jumps from zero
to only $\sim 0.4$ at $T=T_c$ and then increases
with temperature over the range $T_c < T < 4T_c$ \cite{Kaczmarek:2002mc,Dumitru:2003hp}.
As for the potential between heavy quarks,
it was found \cite{Asakawa:2003re} that charmonium states
persist up to $T=1.6T_c$.
The range of temperatures above $T_c$ up to  $(2 \div 4)T_c$ is referred to as the
postconfinement domain,
and strongly interacting matter at these temperatures --- as semi-QGP
\cite{Dumitru:2010mj,Hidaka:2015ima}. It is often considered that color charges are unscreened in `total' QGP,
and partially screened in semi-QGP \cite{Hidaka:2008dr}.

The postconfinement domain was also characterized in terms of
density of Abelian magnetic monopoles \cite{Liao:2006ry,Chernodub:2006gu}.
According to Ref.~\cite{Chernodub:2006gu} the monopoles  are condensed in the confinement phase,
represent dilute gas above $2T_c$
and liquid at $T_c<T<2T_c$. In Ref.~\cite{Liao:2006ry} the relevance of
color-magnetic and color-electric fluctuations was discussed in terms of respective
couplings. The authors also discussed  the screening masses as indicators of the 
(in their terminology) E-M equilibrium point and argued that the temperature
determined by the equality of the electric and magnetic screening masses 
should coincide with temperature determined by respective couplings. Using lattice
results for the electric and magnetic screening masses obtained in $SU(3)$ gluodynamics
\cite{Nakamura:2003pu} they concluded that this temperature is between $1.2T_c$ 
and $1.5T_c$.  

It has long been known \cite{Mitrjushkin:1988vv} that, over the range $T_c<T<3T_c$,
the contributions of the electric and magnetic parts
of the gluon condensate $$\langle G^2_{\mu\nu} \rangle=\langle G^2_{E} \rangle + \langle G^2_{M} \rangle$$
to the energy density and pressure change with
temperature much more rapidly than their sum.
This gives some evidence that an interplay of electric and magnetic degrees of freedom
plays an important role in the dynamics of semi-QGP.

Yet another evidence for this comes from the effective
high-temperature 3D theories \cite{Appelquist:1981vg,Ginsparg:1980ef,Braaten:1995jr},
in which the electric and magnetic degrees of freedom are in fact separated.

For a long time these models failed to reproduce the pressure
of semi-QGP and to predict the transition to the confinement phase
\cite{Kajantie:2002wa, Laine:2005ai}.
This problem was solved with the appearance of the effective
models based on the Polyakov loop \cite{Meisinger:2001cq, Pisarski:2006hz, deForcrand:2008aw},
where large fluctuations of $A_0$ are taken into consideration.
That is, dynamics just above $T_c$  in the 3D effective models
is substantially determined by the electric degrees of freedom.
This resembles the above-mentioned situation with the electric and magnetic condensates
in the postconfinement domain, where the magnetic contribution to the energy density and pressure
only partially compensates the electric one.

In this work we employ both the asymmetry
and the ratio between the transverse and longitudinal propagators
in attempt to set the upper limit on the temperature range where the electric fluctuations dominate.

In \cite{Chernodub:2008kf} the asymmetry was computed for the first time in lattice
$SU(2)$ gluodynamics for a wide range of temperatures
in both confinement and deconfinement phases. It was found that it peaks at the phase
transition and monotonically decreases with increasing temperature in the deconfinement phase.
Furthermore, it was found that the asymmetry crosses
zero at $T\approx 2.2T_c$
and becomes negative at higher temperatures. The existence of this
symmetric point was one of the main results of Ref.~\cite{Chernodub:2008kf}.

In this paper we make a number of improvements in computation of the asymmetry in
comparison with Ref.~\cite{Chernodub:2008kf}. We take care of the finite-volume and
Gribov-copy effects. As a result we demonstrate that the asymmetry is indeed
monotonically decreasing function in the deconfinement phase but it never turns zero.
This result is in a qualitative agreement with the perturbative calculations, see below.

Thus we demonstrate that the asymmetry cannot serve as an indicator of the boundary of the postconfinement
domain. We suggest a new
gluonic quantity to indicate such boundary - the ratio of the magnetic to electric  propagator
at zero momentum which can be related to the ratio of the respective screening masses.

\section{Definitions and simulation details}
\label{sec:definitions}

We study SU(2) lattice gauge theory with the standard Wilson action
\beq
S  = \beta \sum_x\sum_{\mu >\nu}
\left[ 1 -\frac{1}{2}\tr \Bigl(U_{x\mu}U_{x+\mu;\nu}
U_{x+\nu;\mu}^{\dagger}U_{x\nu}^{\dagger} \Bigr)\right], \nonumber
\label{eq:action}
\eeq

\noi where $\beta = 4/g^2$ and $g$ is a bare coupling constant. The
link variables $U_{x\mu} \in SU(2)$ transform  under gauge
transformations $\omega_x$ as follows:

\beq
U_{x\mu} \stackrel{\omega}{\mapsto} U_{x\mu}^{\omega}
= \omega_x^{\dagger} U_{x\mu} \omega_{x+\mu} \; ;
\qquad \omega_x \in SU(2) \; .
\label{eq:gaugetrafo}
\eeq

\noi Our calculations were performed on the asymmetric lattices with
lattice volume $V=N_t\times N_s^3$, where $N_t$ is the number of sites in
the $4th$ direction. The temperature $T$ is given by
\beq
T = \frac{1}{aN_t}~,
\eeq
\noi where $a$ is the lattice spacing.
We employ the standard definition of the lattice gauge vector
potential\footnote{In perturbation theory, $A_{x+\hat{\mu}/2,\mu}$
instead of $A_{x,\mu}$ provides a more adequate designation; $\hat \mu$
is the unit vector in the $\mu$th direction.}
$A_{x,\mu}$ \cite{Mandula:1987rh}:
\beq
A_{x,\mu} = \frac{Z}{2iag}~\Bigl( U_{x\mu}-U_{x\mu}^{\dagger}\Bigr)
\equiv A_{x,\mu}^a \frac{\sigma_a}{2} \,,
\label{eq:a_field}
\eeq
where $Z$ is the renormalization factor,
defined in the text after equation (\ref{eq:DL_DT_zeromom}).

The lattice Landau gauge fixing condition is
\beq
(\nabla^B A)_{x} \equiv {1\over a} \sum_{\mu=1}^4 \left( A_{x,\mu}
- A_{x-a\hat{\mu},\mu} \right)  = 0 \; ,
\label{eq:diff_gaugecondition}
\eeq

\noi which is equivalent to finding an extremum of the gauge functional

\beq
F_U(\omega) = ~\frac{1}{4V}\sum_{x\mu}~\frac{1}{2}~\tr~U^{\omega}_{x\mu} \;,
\label{eq:gaugefunctional}
\eeq

\noi with respect to gauge transformations $\omega_x~$.  After replacing
$U \Rightarrow U^{\omega}$ at the extremum the gauge condition
(\ref{eq:diff_gaugecondition}) is satisfied.

The gluon propagator $D_{\mu\nu}^{ab}(p)$ is defined
as follows:
\beq
D_{\mu\nu}^{ab}(p) = \frac{1}{Va^4}
    \langle \widetilde{A}_{\mu}^a(q) \widetilde{A}_{\nu}^b(-q) \rangle
    \nonumber \\
\label{eq:gluonpropagator}
\eeq

\noi where

\beq
\widetilde A_\mu^b(q) = a^4 \sum_{x} A_{x,\mu}^b
\exp\Big(\ iq(x+{\hat \mu a\over 2}) \Big),
\eeq
$q_i \in (-N_s/2,N_s/2]$ and $ q_4 \in (-N_t/2,N_t/2]$.
The physical momenta $p_\mu$ are defined by relations $ap_{i}=2 \sin{(\pi q_i/N_s)}$,
$ap_{4}=2\sin{(\pi q_4/N_t)}$.

On the asymmetric lattice there are two tensor
structures for the gluon propagator \cite{Kapusta}~:

\beq
D_{\mu\nu}^{ab}(p)=\delta_{ab} \left( P^T_{\mu\nu}(p)D_{T}(p) +
P^L_{\mu\nu}(p)D_{L}(p)\right)\,,
\eeq

\noi where (symmetric) orthogonal projectors $P^{T;L}_{\mu\nu}(p)$
are defined for $p=(\vec{p}\ne 0;~p_4=0)$ as follows

\bea
P^T_{ij}(p)&=&\left(\delta_{ij} - \frac{p_i p_j}{\vec{p}^2} \right),\,
~~~P^T_{\mu 4}(p)=0~;
\\
P^L_{44}(p) &=& 1~;~~P^L_{\mu i}(p) = 0 \,.
\eea

\noi Therefore, two scalar propagators - longitudinal $D_{L}(p)$ and
transverse $D_T(p)$ -  are given by

\bea
D_T(p)&=&\frac{1}{6} \sum_{a=1}^{3}\sum_{i=1}^{3}D_{ii}^{aa}(p)~;
\nonumber \\
D_L(p)&=& \frac{1}{3}\sum_{a=1}^{3} D_{44}^{aa}(p) \,, \nonumber \\
\label{gluonpropagator}
\eea

For $\vec{p} = 0$ they are defined as follows:
\bea\label{eq:DL_DT_zeromom}
D_T(0) &=& \frac{1}{9} \sum_{a=1}^{3} \sum_{i=1}^{3} D^{aa}_{ii}(0)\,,
\nonumber \\
D_L(0) &=& \frac{1}{3}\sum_{a=1}^{3} D^{aa}_{00}(0)\  .
\eea
$D_T(p)$  is associated with the magnetic
sector, $D_L(p)$ -- with the electric sector.

We consider both bare quantities (either labeled by index {\it bare} or without index)
and renormalized quantities (labelled by index {\it MOM}).
The renormalization factor (see equation (\ref{eq:a_field}))
for bare quantities is equal to unity $Z \equiv Z_{bare}=1$.
For renormalized quantities $Z \equiv Z_{MOM}$ and is defined by the requirement
\beq
D_L^{MOM}(p^2=\mu^2)={1\over \mu^2},
\label{renorm_condition}
\eeq
with normalization point $\mu=3$~GeV.

In terms of lattice variables, the asymmetry has the form
\beq
\aasymt= {4a^2 N_t^2\over \beta }
\sum_{b=1}^3\left( \Big\langle A_{x,4}^b A_{x,4}^b \Big\rangle - {1\over 3}\sum_{i=1}^3
 \Big\langle A_{x,i}^b A_{x,i}^b \Big\rangle \right),
\eeq
It can be expressed in terms of the gluon propagators:
\bea\label{eq:average_bare_asymmetry}
&& \aasymt  = {4 N_t \over \beta a^2 N_s^3}
\Big[ 3 (D_L(0) - D_T(0))  \\ \nonumber
&+& \sum_{p\neq 0}\left(
{3|\vec p|^2 \,-\, p_4^2\over p^2}  D_L(p) - 2  D_T(p)\right)\Big] \nonumber
\eea
In the continuum limit, the respective integral is
ultraviolate finite \cite{Chernodub:2008kf,Vercauteren:2010rk};
therefore, no additional renormalization is needed and
this formula holds true for renormalized quantities as well.
Thus the asymmetry $\aasymt$, which is nothing but
the vacuum expectation value of the respective composite operator,
is multiplicatively renormalizable and its renormalization factor
coincides with that of the propagator\footnote{We assume
that $D_L(p)$ and $D_T(p)$ are renormalized with the same factor.}.
\vspace{2mm}

The authors of \cite{Vercauteren:2010rk} obtained one-loop
perturbative estimates of the asymmetry both at high temperatures
\beq\label{eq:HighTempAsymmetry_PT}
\aasym \simeq  {g^2 T^2 \over 4} \left( 1- {g\over 3 \pi}\sqrt{2\over 3} \right)
\eeq
and at low temperatures
\beq
  \aasym \simeq  {g^2 \pi^2\over 10}\;\left(1\, -\, {85\over 522}\,
{g^2\over 16 \pi^2} \right) { T^4 \over M^2} \; ,
\eeq
where
$$
M^2 = -\; {13\over 54}\;  \big\langle A^2 (T=0) \big\rangle\, .
$$

We have generated ensembles of $O(1500)$ independent Monte Carlo
lattice field configurations. Consecutive configurations (considered as
independent) were separated by $100\div 200$ (for $N_s = 24\div 84$)
sweeps, each sweep consisting of one local heatbath update followed
by $N_s/2$ microcanonical updates. In \Tab{tab:statistics} we provide
information about the ensembles used throughout this paper.

\vspace{2mm}

In the gauge fixing procedure we employ the $Z(2)$ transformation
proposed in \cite{Bogolubsky:2005wf}. $Z(2)$ flip in direction $\mu$
consists in flipping all link variables $U_{x\mu}$ attached
and orthogonal to a 3d plane by multiplying them with $-1$.
Such global flips are equivalent to  non-periodic gauge transformations
and represent an exact symmetry of the pure gauge action.
The Polyakov loops in the direction of the chosen links and averaged
over the 3d plane obviously change their sign.  At finite temperature we
apply flips only to directions $\mu=1,2,3$, thus we consider 8 flip sectors.
 In the deconfinement phase,
where the $Z(2)$ symmetry is broken, the $Z(2)$ sector of the Polyakov
loop in the $\mu=4$ direction has to be chosen
since on large enough volumes all lattice configurations belong to the
same sector, i.e. there are no flips between sectors in the Markov chain of
configurations. We choose the sector with positive Polyakov loop.

Following Ref.~\cite{Bornyakov:2009ug} in what follows we call the
combined gauge fixing algorithm employing simulated annealing (SA) algorithm
(with finalizing overrelaxation) and $Z(2)$ flips for space directions the `FSA' algorithm.
We generated $n_{copy}= 1$ to 3 gauge copies per flip--sector
each time starting from a random gauge transformation of
the Monte Carlo configuration, obtaining in this way $N_{copy} = 8 n_{copy}$ Landau-gauge
fixed copies for every configuration.  We take the copy with maximal value of the functional
(\ref{eq:gaugefunctional}) as our best estimator of the global maximum
and denote it as best (``\bc'') copy.  In order to demonstrate the Gribov
copy effect we compare with the results obtained from the randomly chosen
first (``\fc'') copy and with the 'worst' copy ("{\it wc}"), i.e. copy with the lowest value of the
gauge functional.

To suppress 'geometrical' lattice artifacts, we apply the
``$\alpha$-cut'' \cite{Nakagawa:2009zf}, i.e.  $~\pi q_i/N_s < \alpha~$,
for every component, in order to keep close to a linear behavior
of the lattice momenta $p_i \approx (2 \pi q_i)/(aN_s), ~~q_i \in
(-N_s/2,N_s/2]$.  We have chosen $\alpha=0.5$. Obviously, this cut
influences large momenta only.
We did not employ the {\it cylinder cut} in this work.

\section{$A^2$ asymmetry in the deconfinement phase}

The asymmetry was introduced and studied numerically in \cite{Chernodub:2008kf}
in a rather wide range of temperatures ($0.4~T_c < T < 6~T_c$).
The computations were made on the lattices
$16^3 \times 4$, $24^3 \times 6$, and $32^3 \times 8$.
A nontrivial temperature dependence was obtained.
In particular, it was found that the asymmetry is positive at
$T<2.21(5) T_c$ and negative at $T > 2.21(5) T_c$. This observation
was considered as an indication that at
high temperatures magnetic fluctuations 
begin to dominate. Comparing data for three lattice spacings the authors
concluded that finite lattice spacing effects are small even for $N_t=4$.
This allows us to assume that our results obtained on lattices with $N_t=8$
are also free of substantial finite lattice spacing effects.

Let us note that in \cite{Chernodub:2008kf} (as well as in this work)
the temperature was changed by variation of the lattice spacing for fixed $N_t$.
In Ref.~\cite{Chernodub:2008kf} the finite volume effects were not checked although
the spatial lattice size was decreasing with increasing temperature and
at the highest temperature $T=6~T_c$ it was as small as $L \equiv aN_s = 0.44$~fm
with the corresponding minimal momentum $p_{min}\simeq 2.8$~GeV.
In this work we carefully study the finite volume effects using lattices up
to $L=3$~fm (the detailed information on lattices used in this work is given in
Table~\ref{tab:statistics}).
Furthermore, we  use $Z_2$ flips which help to reduce finite volume effects
as was found in Ref.~\cite{Bogolubsky:2007bw}.
Here we again show that the effect of flip sectors is very substantial
on small volumes. We then demonstrate that taking care about the finite volume effects
dramatically changes some of the conclusions made in \cite{Chernodub:2008kf}.

First, we want to reproduce the results obtained in \cite{Chernodub:2008kf}
at high temperatures ($2 < T/T_c < 6$).

\begin{figure}[tbh]
\includegraphics[width=7.5cm]{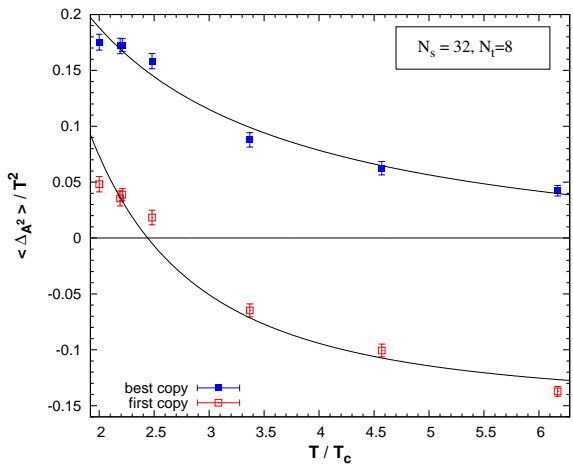}
\caption{Asymmetry on lattices $32^3 \times 8$ as function of temperature.
Lower data set shows our results for first copy
and should be compared with that obtained in
\cite{Chernodub:2008kf}. Upper data set was
obtained with the FSA algorithm (best copy). The curves show
results of the fits to eq.~(\ref{eq:asym_vs_xi_power})
(for first copy) and to eq.~(\ref{eq:fit_bc_FSA}) (for
best copy).
}
\label{fig:A2_32_vs_T}
\end{figure}

In Fig.\ref{fig:A2_32_vs_T} we show our results for $\aasymt$ obtained on
lattices $32^3\times 8$ used in \cite{Chernodub:2008kf}. Lower data set
shows our results for the first copy ({\it fc}).
These results are to be compared with those obtained in
\cite{Chernodub:2008kf}. Upper data set corresponds to the best copy ({\it bc}).
One can see that two data sets differ dramatically and this difference grows with
temperature.

To make explicit comparison with \cite{Chernodub:2008kf}, we fit
data points corresponding to {\it fc} copy to the function
\beq\label{eq:asym_vs_xi_power}
\hspace*{-3mm}
\aasymt =  b_0 + {b_2 \over \xi^2}\; ,
\eeq
used in \cite{Chernodub:2008kf}; here and below $\xi=T/T_c$.
The parameters obtained in our fit,
\beq
b_{0}=\;-\;0.15(1), \ b_{2}=\big[ 0.946(37) \big]^2 \ , \quad
\eeq
agree well with those found in  \cite{Chernodub:2008kf}:
\beq
b_{0}=\;-\;0.164(4), \ b_{2}=\big[ 0.894(14) \big]^2\ .
\eeq

Our value $\xi = 2.44(13)$ at which $\aasymt_{fc} =0$
is only a little higher than the respective value $\xi=2.21(5)$ from
\cite{Chernodub:2008kf}. We conclude that
our values of $\aasymt_{fc}$ come close to
the values of the asymmetry obtained in \cite{Chernodub:2008kf}.

Now we turn to the upper data set.
It differs significantly from both the {\it fc} data set
and the results of \cite{Chernodub:2008kf}.
The main qualitative difference is that $\aasymt_{bc}$
does not cross zero within the range of temperatures under study.
Since the difference between the two procedures employed
to obtain these two data sets consists in the use of flips,
we attribute the observed difference to the flip effects.
As was shown in \cite{Bogolubsky:2007bw} the use of flips
substantially reduces finite volume effects,
thus we expect that the observed difference
increases with a decrease of the lattice size.

In the case of {\it bc} data set the best fit is provided by
the fit function
\beq\label{eq:fit_bc_FSA}
\aasymt_{bc} \simeq b_{0} + {b_{1}\over \xi },
\eeq
with parameters
\beq
b_{0}=-0.03(1), \ b_{1}= 0.44(3)  \ , \quad
\eeq
and $\ds {\chi^2\over N_{dof}} = 2.22$. 
Even if $\aasymt_{bc}$ at $N_s=32$ becomes negative,
this occurs at temperatures much greater than the upper
limit of the range under our consideration.

\begin{figure}[t]
\vspace*{.7cm}
\includegraphics[width=8cm]{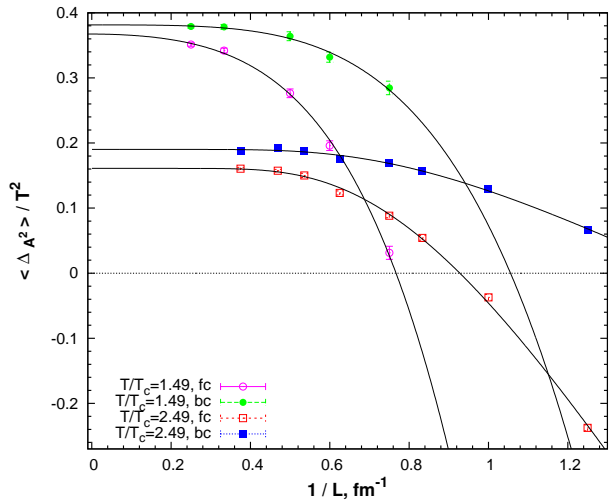}
\caption{Lattice size dependence of the asymmetry at $T/T_c=1.49$ and $2.49$.
Empty circles show {\it fc} results, full circles - {\it bc} results. Curves
show results of the fit to eq.~(\ref{fit_expon}).
}
\label{fig:A2_vs_L_norm_fit}
\end{figure}
Now we proceed to the study of the finite-volume dependence of the asymmetry
and infinite volume extrapolation.
In Fig.\ref{fig:A2_vs_L_norm_fit} we show
lattice-size dependence of the asymmetry
at $T/T_c=1.49$ and $T/T_c=2.49$.
Empty symbols show the {\it fc} results, filled symbols - the {\it bc} results.
As is seen in Fig.\ref{fig:A2_vs_L_norm_fit}, the volume dependence
of $\aasymt_{fc} $ is very significant and at $L \simeq 1.3$~fm for $T/T_c=1.49$
($L \simeq 1.1$~fm for $T/T_c=2.49$)
it even changes sign. As expected the finite-size effects for $\aasymt_{bc}$
are much smaller and this is due to flips.
Still the data indicate that to reduce finite-size effects below $3 \% $
one needs the minimal lattice size about 2.5~fm for both $T/T_c=1.49$ and 
$T/T_c=2.49$. 

\begin{table}[tbh]
\begin{center}
\vspace*{0.2cm}
\begin{tabular}{|c|c|c|c|c|c|c|c|c|} \hline
     &  Gauge      &       &                      &         &  \\
$\xi$&  fixing     & ~~$\aasymt_\infty^{pol}$~~ & ~$\sqrt{c_2}$,~fm~ & ~$\sqrt[4]{c_4}$,~fm~ 
 &~$\ds {\chi^2\over N_{dof}}$~  \\
     & ~algorithm~ &       &                      &        &                      \\
   \hline\hline
1.49 & ~$bc$~  &  0.3828(49) & $0.1^{+0.3}_{-0.1}$& 0.70(8) & 0.76    \\
1.49 & ~$fc$~  &  0.3674(64) &  0.428(90)         & 0.92(4) & 1.61   \\
2.49 & ~$bc$~  &  0.1965(37) &  0.179(40 )        & 0.43(3) & 1.99    \\
2.49 & ~$fc$~  &  0.1834(48) &  0.336(27)         & 0.56(2) & 3.41   \\
\hline\hline
\end{tabular}
\end{center}
\caption{Results of fitting of the asymmetry to polynomial fit eq.~(\ref{fit_polynom}).
}
\label{tab:asym_vs_L_pow}
\end{table}
To compute the asymmetry in the infinite volume limit $\aasymt_\infty$
we begin with the polynomial fit of the type
\beq
\aasymt (L) = \aasymt_\infty^{pol} - {c_2\over L^2} -  {c_4\over L^4}\; ,
\label{fit_polynom}
\eeq
the results are shown in Table~\ref{tab:asym_vs_L_pow}.

To estimate systematic errors due to choice of the fitting function, we also fitted the data
to the fit function
\beq
\aasymt (L) \simeq \aasymt_\infty^{exp} - c \exp \Big(-\; L/L_0 \Big)\;
\label{fit_expon}
\eeq
which provides even better quality.
The results of this fit are presented in Table~\ref{tab:asym_vs_L_exp}. One can see that
the values of $\aasymt_\infty^{pol}$ and $\aasymt_\infty^{exp}$ 
agree within statistical error bars. This implies
that the systematic error is of the same order as the statistical one.

\begin{table}[tbh]
\begin{center}
\vspace*{0.2cm}
\begin{tabular}{|c|c|c|c|c|c|c|c|c|} \hline
     &  Gauge      &       &       &           &        \\
$\xi$&  fixing     & ~~$\aasymt_\infty^{exp}$~~&~~$c$~~& ~$L_0$~(fm)~ & ~${\chi^2\over N_{dof}}$~ \\
     & ~algorithm~ &       &       &           &    \\
   \hline\hline
1.49 & ~$bc$~  & 0.380(2) & 1.7(1.0) & 0.41(5)  & 0.34   \\
1.49 & ~$fc$~  & 0.352(1) & 4.7(1.0) & 0.47(8)  & 0.06  \\
2.49 & ~$bc$~  & 0.190(2) & 1.7(5) & 0.31(3)    & 1.71   \\
2.49 & ~$fc$~  & 0.161(2) & 5.6(5) & 0.31(1)    & 2.60  \\
\hline\hline
\end{tabular}
\end{center}
\caption{ Results of fitting of the asymmetry to the exponential fit eq.~(\ref{fit_expon}).
}
\label{tab:asym_vs_L_exp}
\end{table}

We have also checked  finite size effects for the dimension 2
electric and magnetic condensates
separately. We found that the electric condensate is constant within error  bars,
whereas the magnetic one decreases  with increasing volume. Thus the finite size
effects in the asymmetry are due to volume dependence of the magnetic condensate.

\begin {figure}[tbh]
\includegraphics[width=8cm]{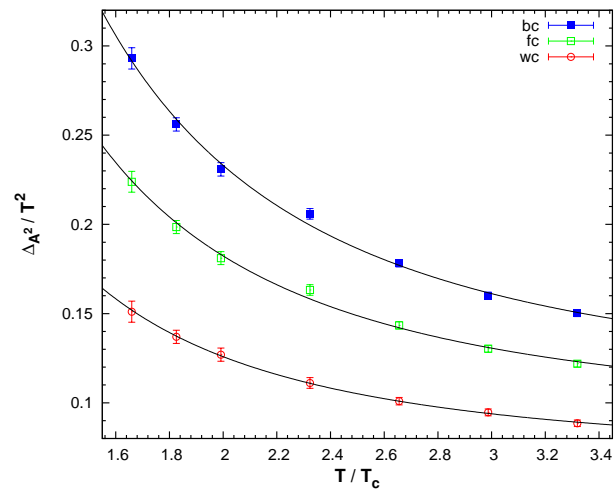}
\caption{Asymmetry as function of the temperature for
the fixed lattice size $L=2$~fm.
The curves show the fit function (\ref{eq:asym_vs_xi_power})
at the respective values of the parameters.}
\label{fig:A2_vs_T_2fm_08_lar}
\end{figure}

Next we consider the temperature dependence of the condensate.
The results for the asymmetry in the case of fixed lattice size $L=2$~fm
and varying temperature are shown in Fig.\ref{fig:A2_vs_T_2fm_08_lar}.
We found that a good fit is provided by the function (\ref{eq:asym_vs_xi_power}).
The respective fit parameters are shown in Table~\ref{tab:asym_vs_T_fit00}.
It should be noted that this fit function
works at $T> 1.6 T_c$; at smaller temperatures
terms of the order $T_c^4 / T^4$ {\it etc} are necessary.

In Fig.\ref{fig:A2_vs_T_2fm_08_lar} we also show
the results for the 'worst' copy  which was first
introduced in \cite{Bornyakov:2013ysa}. For a given configuration
the worst copy is defined as a gauge copy with the
lowest value of the gauge fixing functional.
The worst copy results are to demonstrate that the Gribov copy
effects within first Gribov horizon are substantially
stronger than the difference between our first copy and best copy.

Using the fit function (\ref{eq:asym_vs_xi_power}), we find that
the asymmetry is positive at all temperatures. This is in agreement
with the perturbative result (\ref{eq:HighTempAsymmetry_PT}).

\begin{table}[tbh]
\begin{center}
\vspace*{0.2cm}
\begin{tabular}{|c|c|c|c|} \hline
  Gauge     &             &           &                             \\
  fixing    & ~~$b_0$~~   & ~~$b_2$~~ & ~${\chi^2\over N_{dof}}$~ \\
~algorithm~ &             &           &                             \\
            &             &           & $\xi>1.65$        \\
            &             &           &                            \\
   \hline\hline
~$bc$~  &  0.1036(27) & 0.517(16) &    1.40        \\
~$fc$~       &  0.0893(22) & 0.372(13) &    0.92       \\
~$wc$~  &  0.0682(5)  & 0.231(3)  &    0.05        \\
\hline\hline
\end{tabular}
\end{center}
\caption{$b_0$ and $b_2$ are the parameters of the
fit (\ref{eq:asym_vs_xi_power}) performed over the range
$1.65 < \xi < 3.32$ for asymmetry computed at fixed lattice size $L=2$fm.
}
\label{tab:asym_vs_T_fit00}
\end{table}

The renormalized asymmetry $\aasymt^{MOM}=Z_{MOM} \aasymt$ where $Z_{MOM}$
is determined in (\ref{renorm_condition}) can also be fitted by the function
(\ref{eq:asym_vs_xi_power}) with result
\beq
\aasymt_{bc}^{MOM} = 0.0602(18)\;+\;{0.268(11)\over \xi^2},\qquad {\chi^2 \over N_{dof}}=0.91.
\eeq
The fit was performed over the same range of temperatures.

However, the fit function (\ref{eq:asym_vs_xi_power}) disagrees
with the perturbative result (\ref{eq:HighTempAsymmetry_PT}) in the limit 
of infinite
temperature where perturbation symmetry is believed to be valid.
For this reason, we also fit the data to the function (motivated by (\ref{eq:HighTempAsymmetry_PT}))
\beq\label{eq:pert_2}
\aasymt \simeq  {z g^2(T) \over 4} \left(1 - {g(T)\over 3 \pi}\sqrt{2\over 3} \right)\; ,
\eeq
where the running coupling is taken in the two-loop approximation,
\beq \label{eq:two-loop_coupling}
{1\over g^2(T)} = {1\over 4\pi^2}\;
\left({11\over 6} \ln \Big( {T^2\over \Lambda^2} \Big)
 + {17\over 11} \ln\ln \Big( {T^2\over \Lambda^2} \Big) \right)\; ,
\eeq
$z$ and $\Lambda$ are the fit parameters.

\begin{figure}[tbh]
\includegraphics[width=8cm]{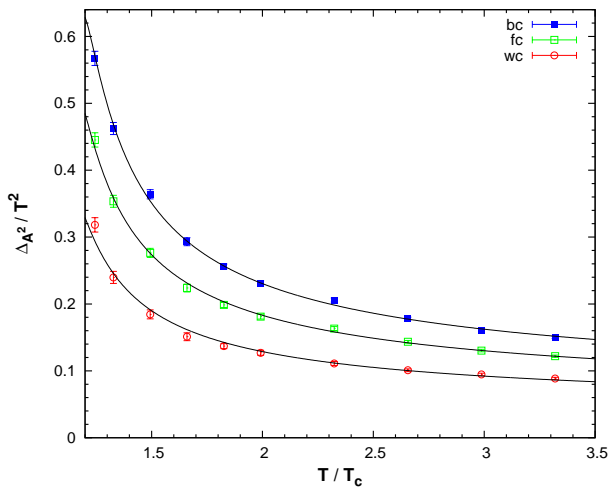}
\caption{Same as in Fig.~\ref{fig:A2_vs_T_2fm_08_lar}
but $\aasymt $ is fitted to the function (\ref{eq:pert_2}); also
wider range of temperatures is shown.
}
\label{fig:asym_vs_T_pert_3}
\end{figure}

\begin{table}[tbh]
\begin{center}
\vspace*{0.2cm}
\begin{tabular}{|c|c|c|c|} \hline
  Gauge     &             &            &                              \\
  fixing    & ~~$z$~~   & ~~$\Lambda/T_c$~~& ~${\chi^2\over N_{dof}}$~ \\
~algorithm~ &             &            &                    \\
            &             &            &  $\xi>1.24$        \\
            &             &            &                        \\
   \hline\hline
~$bc$~  & 0.1284(14) & 0.845(7)   &    1.50      \\
~$fc$~  & 0.1045(13) & 0.826(7)   &    1.12      \\
~$wc$~  & 0.0749(17) & 0.811(14)  &    1.93      \\
\hline\hline
\end{tabular}
\end{center}
\caption{Parameters of the fit of the asymmetry to eq.~(\ref{eq:pert_2}). }
\label{tab:asym_vs_T_fit03}
\end{table}

The results of the fit over the range $1.24 < \xi < 3.32$
are shown in Table~\ref{tab:asym_vs_T_fit03}. Note that it works over
wider range of temperatures than the fit (\ref{eq:asym_vs_xi_power}).

Therefore, we arrive at a good agreement with perturbation theory
modulo the normalization factor of the propagator.
In order to make quantitative comparison with the perturbative result,
we should use the same normalization condition; however,
the $\overline {MS}$ scheme employed in \cite{Vercauteren:2010rk}
runs into difficulties beyond perturbation theory.

Thus we infer that, contrary to the conclusions of
Ref.~\cite{Chernodub:2008kf},
the asymmetry never crosses zero in the deconfining
phase. Accordingly it cannot indicate the boundary between two regions
of the deconfining phase, whose existence was discussed in
\cite{Liao:2006ry,Chernodub:2006gu,Liao:2008jg,Dumitru:2010mj,Chernodub:2008vn}.

\section{Ratio $D_T(0)/D_L(0)$}

It is natural to expect that the existence of the postconfinement region
is explained by the contribution of low-momentum,
i.e. nonperturbative, modes of the gauge field.
This motivates us to consider in this work the ratio of
the magnetic to electric propagator at zero momentum
as a possible indicator of the boundary of the postconfinement region.

Similar ratio of electric and magnetic screening masses
was computed in \cite{Heller:1997nqa}.
These masses were evaluated in \cite{Heller:1997nqa} (see analogous computation
in $SU(3)$ gluodynamics in \cite{Nakamura:2003pu}) in a
renormalization-invariant way, by long-distance behavior of
the gluon propagators:
\bea
\tilde D_L(p_\perp = 0,x_3) &\sim& \exp(-m_e |x_3|),  \\
\tilde D_T(p_\perp = 0,x_3) &\sim& \exp(-m_m |x_3|),\,\,|x_3|\to \infty  \nonumber
\eea
where $\tilde D_L(p_\perp,x_3)$ and $\tilde D_T(p_\perp,x_3)$
are, respectively, the Fourier transforms of $D_L(p)$ and $D_T(p)$ in the third
component of the momentum.
There are different views on gauge invariance of these masses
--- even in the framework of perturbation theory:
the authors of \cite{Heller:1997nqa} consider them gauge-independent,
whereas the authors of \cite{Kapusta} cast some doubt on
both their gauge-invariance and physical meaning.

In the leading-order perturbation theory
$\ds m_e = \sqrt{\frac{2}{3}} g(T) T$ in the $SU(2)$ case, 
whereas for $m_m$ (which is of nonperturbative nature) 
the behavior  $~g^2(T) T$ is conjectured.
The authors of \cite{Heller:1997nqa} obtained the data for $T>2T_c$ and employed
fit formula
\beq\label{mass_ratio}
{m_e^2(T) \over m_m^2(T)} = {C\over g^2(T)}.
\eeq
Let us note that the authors of \cite{Heller:1997nqa} did not
take care of finite size effects. In their study the lattice size was decreasing
with an increase of the temperature similar to Ref.~\cite{Chernodub:2008kf}.
Over the temperature range explored in our study, their lattice size
decreased from 2~fm down to 0.8~fm.

In the present work we consider the ratio $\ds r(T)={D_T(0)\over D_L(0)}$
instead of $\ds {m_e^2\over m_m^2}$. Our arguments are as follows.
It was shown in
\cite{Bornyakov:2010nc} that the low momentum behavior of $D_L(p)$ 
is compatible with pole behavior and renormalized $D_L(0)$ can be considered
as inverse electric mass squared. However, the low momentum
behavior of $D_T(p)$ is definitely different from the pole behavior.
It has a maximum at
nonzero momentum $p_0 \sim 0.4 \div 0.5$~GeV, see, e.g. Fig.~8 in
\cite{Bornyakov:2010nc}. Still $D_T(0)$
characterizes the strength of $D_T(p)$ at low momentum\footnote{
Note that Linde \cite{Linde:1980ts} related the magnetic mass to transverse
gluon propagator at zero momentum.}
We assume that the temperature $T_p$ satisfying relation 
\beq
r(T_p)=1\ ;
\label{eq:Tp_definition}
\eeq
determines the boundary of the postconfinement region. This is not a phase transition
thus the boundary is not characterized by definite value of the temperature but rather by
a range of temperature values.

\begin{figure}[tbh]
\includegraphics[width=8cm]{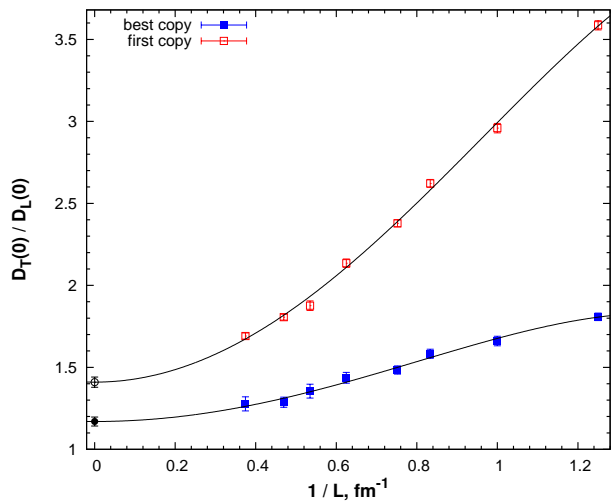}
\caption{Lattice size dependence of the ratio $r(T)$ for first copies (empty symbols)
and best copies (filled symbols) at $T/T_c=2.49$. }
\label{fig:DT_to_DL_vs_vol_fit}
\end{figure}

We start from the study of the finite size effects for $r(T)$. In
Fig.~\ref{fig:DT_to_DL_vs_vol_fit} we show the ratio as the function of the inverse lattice
size $1/L$ for $T/T_c=2.49$ for first copies (empty symbols) and best copies (filled symbols).
The difference between two data sets is huge on small volumes and decreases  with increasing
volume. As in the case of the asymmetry (see Fig.~\ref{fig:A2_vs_L_norm_fit}) this difference is
due to application of the flip procedure. For best copies the finite size effects are much
smaller than for the first copies but they are still sizable up to lattice size $L$ about 2~fm.
For  $L=3$fm lattice  finite size correction is small and comparable to 
statistical error.
We fit the lattice size dependence of $r(T)$ to the polynomial fit as in the case of asymmetry
\beq
r(T,L)=r(T,\infty) + {r_2(T)\over L^2} + {r_4(T)\over L^4}\ .
\eeq
The result is as follows:
\bea
\hspace*{-8mm}{bc}: &&  r(2.49~T_c, \infty)= 1.170(27), \  {\chi^2\over N_{dof}}=0.57\; ; \\ \nonumber
\hspace*{-8mm}{fc}: &&  r(2.49~T_c, \infty)= 1.410(31), \ {\chi^2\over N_{dof}}=2.12\;. \nonumber
\eea
Results of the fits are presented in Fig.~\ref{fig:DT_to_DL_vs_vol_fit}. The fits predict  that the
difference between $r(T)$ values obtained via two gauge fixing procedures survives in the
infinite volume limit.

\begin{figure}[tbh]
\includegraphics[width=8cm]{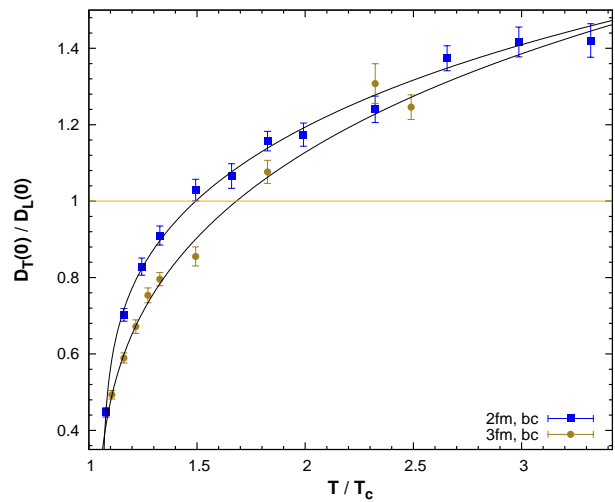}
\caption{Ratio $r(T)$ for best copies on lattices with fixed sizes $L=2$ fm 
and $L=3$ fm  versus temperature. Fits to eq.~\ref{fit_ratio} are also shown.}
\label{fig:DT_to_DL_vs_T}
\end{figure}

The best copy ratio $r(T)$ for two fixed lattice sizes $L=2$ fm and $L=3$ fm  versus temperature is plotted in
Fig.\ref{fig:DT_to_DL_vs_T}. Note that the point on $L=3$ fm lattice at the largest temperature $T=2.49 T_c$ was
obtained by the extrapolation shown in Fig.\ref{fig:DT_to_DL_vs_vol_fit}. 
The data presented in this figure indicate that the finite size effect
is definitely nonzero up to $T/T_c=1.8$ and might disappear at higher temperatures.

We fit the $T$ dependence of $r(T)$ to the function
\beq \label{fit_ratio}
 r(T) \simeq r_0 + {r_1\over g^2(T)}\; ,
\eeq
inspired by eq.~(\ref{mass_ratio}).
Here $r_0,r_1,\Lambda/T_c$ are the fit parameters.
This fit formula works surprisingly well at
$T>1.08 T_c$ as can be seen from Fig.\ref{fig:DT_to_DL_vs_T} and
Table~\ref{tab:DT_to_DL_vs_T_fit1}.

\begin{table}[tbh]
\begin{center}
\vspace*{0.2cm}
\begin{tabular}{|c|c|c|c|c|c|} \hline
Lattice  &            &           &                   &             & \\[-2mm]
 size   & $r_0$~~ & ~~$r_1$~~ & ~$\Lambda / T_c$~ & ~$T_p/T_c$~ & ~$\ds {\chi^2\over N_{dof}}$~  \\[-2mm]
        &           &           &                   &             &  \\
   \hline\hline
2 fm &   0.94(1)  & 3.78(12) & 1.060(3) & 1.494(30) & 0.64    \\
3 fm &  0.79(3)  & 4.59(37) & 1.02(2)  & 1.68(12)  & 1.42    \\
\hline\hline
\end{tabular}
\end{center}
\caption{Parameters of the fit of the ratio $r(T)$ for best copies on lattices with $L=2$ fm and $L=3$ fm  to fit function
(\ref{fit_ratio}).
}
\label{tab:DT_to_DL_vs_T_fit1}
\end{table}

\begin{figure}[tbh]
\includegraphics[width=8cm]{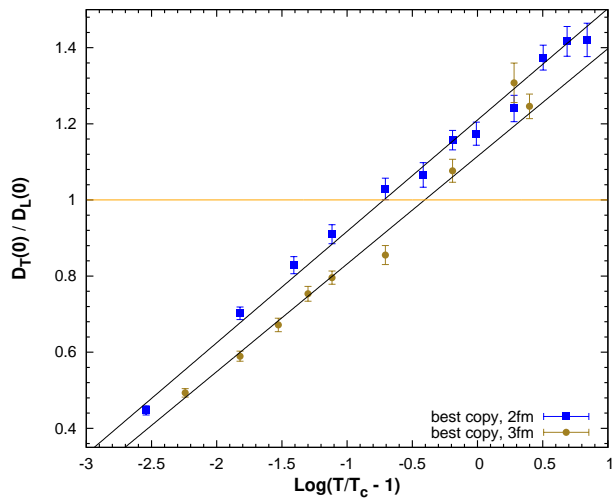}
\caption{Ratio $r(T)$ for best copies on lattices with fixed sizes $L=2$ fm 
and $L=3$ fm  versus $\log(T/T_c-1)$ with fits to eq.~\ref{eq:simple_log_fit}.}
\label{fig:DT_to_DL_fit_aux}
\end{figure}

It should be noted that the fit formula
(\ref{fit_ratio}) works well for all $T>\Lambda$ - even in the case when
the coupling $g^2(T)$ (\ref{eq:two-loop_coupling}) becomes negative at
$T/T_c$ below 1.4.
However, this motivates us to employ yet another fit function
\beq
r(T) \simeq R_0 + R_1 \ln \left( {T\over T_c} -1\right)
\label{eq:simple_log_fit}
\eeq
The results of this fit are shown in Fig.\ref{fig:DT_to_DL_fit_aux}
and in Table~\ref{tab:DT_to_DL_vs_T_fit2}. It is clearly seen that
this simple logarithmic fit function works also  well.
The slope is independent of the volume, whereas the intercept slowly
decreases with an increase of lattice size. We have every reason to consider
$L=3$fm as a good approximation to the infinite-volume limit (cf. 
Fig.~\ref{fig:DT_to_DL_vs_vol_fit}).
The values of $T_p/T_c$ obtained with two fits agree within error bars.

The fit function (\ref{eq:simple_log_fit}) can be rearranged to the form
\beq
r(T) \simeq R_1 \ln \left( {T - T_c\over Q}\right),
\label{eq:simple_log_fit_new}
\eeq
where $Q$ sets the scale over the postconfinement domain.
We find that $Q\approx 230$~MeV, which comes close to the quantity
$m=201(8)$~MeV \cite{Chernodub:2008kf} that sets the temperature
scale for the asymmetry at low temperatures in the confinement phase.

\begin{table}[tbh]
\begin{center}
\vspace*{0.2cm}
\begin{tabular}{|c|c|c|c|c|c|} \hline
Lattice &             &           &              & \\[-2mm]
 size   &   ~~$R_0$~~ & ~~$R_1$~~ &  ~$T_p/T_c$~ & ~$\ds {\chi^2\over N_{dof}}$~  \\[-2mm]
        &            &           &               &  \\
   \hline\hline
2 fm &   1.21(1)  & 0.293(6) &  1.488(13) & 1.35    \\
3 fm &  1.115(15) & 0.283(9) &  1.667(27) & 1.92    \\
\hline\hline
\end{tabular}
\end{center}
\caption{Parameters of the fit of the ratio $r(T)$ for best copies on lattices with $L=2$ fm and $L=3$ fm  to fit function
(\ref{eq:simple_log_fit}).
}
\label{tab:DT_to_DL_vs_T_fit2}
\end{table}
The difference between the parameters in 
Tables \ref{tab:DT_to_DL_vs_T_fit1} and \ref{tab:DT_to_DL_vs_T_fit2}
gives an estimate of the systematic error in determination of $T_p$
defined by the formula (\ref{eq:Tp_definition}).

The fit functions (\ref{fit_ratio}) and (\ref{eq:simple_log_fit})
imply that the ratio $D_T(0)/D_L(0)$ goes to infinity
in the infinite temperature limit. Results at higher temperatures are needed to confirm
this prediction.

\section{Conclusions}

We presented results of the study of the asymmetry $\aasymt$ and the 
ratio $D_T(0)/D_L(0)$
in lattice $SU(2)$ gluodynamics  on lattices with varying spatial size $N_s$
in the range of temperatures above $T_c$ up to $3.3 T_c$. Our findings can be summarized as
follows:

\begin{itemize}

\item In contrast to conclusions made in \cite{Chernodub:2008kf}
 the asymmetry is positive at all temperatures under consideration 
 and its high-temperature 
behavior agrees with perturbation theory.  The data can be fitted to function
motivated by the perturbation theory down to temperatures as low as $1.25 T_c$.
The asymmetry cannot be used as indicator of the postconfinement domain boundary. 
 \item A good indicator of the
 boundary of the postconfinement domain is
 provided by $D_T(0)/D_L(0)$ rather than by the asymmetry $\aasymt$.
 The transition temperature $T_p$ defined by the condition $D_T(0)/D_L(0)=1$
 slightly increases with increasing volume.
 At $L=3fm$, which is close to the infinite-volume limit,
 $T_p = 1.68(12) T_c$.
 \item  In the range of temperatures under study in this work the effect of flip sectors
is substantial at $L\simeq 2$~fm
 and crucial at $L<1$~fm.  In the latter case,
 it dramatically changes the behavior of both the asymmetry and ratio $D_T(0)/D_L(0)$.
 \item Finite-volume effects
 are  significant on lattices with $L < 2$~fm  within our range of temperatures and
decrease with increasing temperature.
 \item The temperature dependence of the ratio $D_T(0)/D_L(0)$ 
 can well be fitted by both the perturbatively motivated function (\ref{fit_ratio}) 
 and the linear function of $\ln(T-T_c)$ (\ref{eq:simple_log_fit_new})
 over the range $1.08T_c < T < 3.32T_c$.
 
\end{itemize}

\acknowledgments{Computer simulations were performed on the IHEP (Protvino)
Central Linux Cluster, ITEP (Moscow) Linux Cluster, MSU 'Lomonosov' supercomputer.
The work was supported by the Russian Foundation for Basic Research, grant no.16-02-01146~A.}


\begin{table*}[h]
{\large\bf Appendix: Table of statistics}
\begin{center}
\vspace*{0.5cm}
\begin{tabular}{|c|c|c|c|c|c|c|c|} \hline
$~\beta$&~$N_s$~& $L$~(fm)&$a^{-1}$~(GeV)&$T$~(MeV)& $~T/T_c~$ &$n_{copy}$& ~$n_{meas}$~ \\ \hline\hline
2.5574 &   28  &   2.00 &    2.7622    &   345.3  & 1.162 & 1    &  1780  \\
2.5792 &   30  &   2.00 &    2.9595    &   369.4  & 1.245 & 1    &  1246  \\
2.5996 &   32  &   2.00 &    3.1568    &   394.6  & 1.328 & 1    &  1068  \\
2.6370 &   36  &   2.00 &    3.5514    &   443.9  & 1.494 & 1    &  1157  \\
2.6706 &   40  &   2.00 &    3.9460    &   493.3  & 1.660 & 1    &   979  \\
2.7011 &   44  &   2.00 &    4.3406    &   542.6  & 1.826 & 1    &  1744  \\
2.7290 &   48  &   2.00 &    4.7352    &   591.9  & 1.992 & 1    &  1246  \\
2.7788 &   56  &   2.00 &    5.5244    &  690.6   & 2.324 & 1    &  1068  \\
2.8221 &   64  &   2.00 &    6.3136    &  789.2  & 2.656 & 2    &  1467  \\
2.8604 &   72  &   2.00 &    7.1028    &  887.9  & 2.987 & 2    &  1132  \\
2.8949 &   80  &   2.00 &    7.8920    &  986.5  & 3.319 & 3    &   927  \\ \hline
2.8011 &   24  &   0.80 &    5.919     &  739.9   & 2.490 & 1    &  3560  \\
2.8011 &   36  &   1.20 &    5.919    & 739.9   & 2.490 & 1    &  2816 \\
2.8011 &   40  &   1.33 &    5.919    & 739.9   & 2.490 & 1    &  2848 \\
2.8011 &   48  &   1.60 &    5.919    & 739.9   & 2.490 & 1    &   801 \\
2.8011 &   56  &   1.87 &    5.919    & 739.9   & 2.490 & 1    &   880 \\
2.8011 &   64  &   2.13 &    5.919    & 739.9   & 2.490 & 2    &   693 \\
2.8011 &   80  &   2.67 &    5.919    & 739.9   & 2.490 & 3    &   740 \\ \hline
2.7310 &   32  &   1.33 &    4.764    &  595.5  & 2.00  & 1    &  1068 \\
2.7600 &   32  &   1.21 &    5.213    &  651.6  & 2.19  & 1    &  1068 \\
2.7630 &   32  &   1.20 &    5.261    &  657.6  & 2.21  & 1    &  1691 \\
2.8000 &   32  &   1.07 &    5.899    &  773.4  & 2.48  & 1    &  1068 \\
2.9000 &   32  &   0.79 &    8.016    &  1002   & 3.37  & 1    &  1068 \\
3.0000 &   32  &   0.58 &    10.86    &  1357   & 4.57  & 1    &  1068 \\
3.1000 &   32  &   0.43 &    14.68    &  1835   & 6.17  & 1    &  1780 \\ \hline
2.5421 &   40  &   3.00 &   2.6307    &  328.8  & 1.106  & 1    &  1758 \\
2.5574 &   42  &   3.00 &   2.7622    &  345.3  & 1.162  & 1    &  1780 \\
2.5721 &   44  &   3.00 &   2.8937    &  361.7  & 1.106  & 1    &  1273 \\
2.5861 &   46  &   3.00 &   3.0253    &  378.2  & 1.106  & 1    &  1412 \\
2.5996 &   48  &   3.00 &   3.1568    &  394.6  & 1.328  & 1    &  1780 \\
2.7011 &   66  &   3.00 &   4.3406    &  542.6  & 1.826  & 2    &  1011 \\  \hline
2.6370 &  24  & 1.33 &    3.5514   &  443.9  & 1.494 & 1   &  1709 \\ 
2.6370 &  30  & 1.67 &    3.5514   &  443.9  & 1.494 & 1   &  1660 \\ 
2.6370 &  36  & 2.00 &    3.5514   &  443.9  & 1.494 & 1   &  1157 \\
2.6370 &  48  & 1.67 &    3.5514   &  443.9  & 1.494 & 1   &  1780 \\ 
2.6370 &  54  & 3.00 &    3.5514   &  443.9  & 1.494 & 1   &  1068 \\ 
2.6370 &  72  & 4.00 &    3.5514   &  443.9  & 1.494 & 2   &  974 \\  \hline

 \hline\hline
\end{tabular}
\end{center}
\caption{Values of $\beta$, lattice sizes, temperatures, number of
measurements and number of gauge copies used throughout this paper.
To fix the scale we take $\sqrt{\sigma}=440$ MeV.
}
\label{tab:statistics}
\end{table*}

\end{document}